\documentclass[12pt,a4paper]{article}
\usepackage{a4wide,cite}
\usepackage{amsmath,amssymb}
\usepackage{latexsym,graphicx,epsfig}

\def\lsim{\mathrel{\rlap{\raise 2.5pt \hbox{$<$}}\lower 2.5pt\hbox{$\sim$}}}
\def\gsim{\mathrel{\rlap{\raise 2.5pt \hbox{$>$}}\lower 2.5pt\hbox{$\sim$}}}

\allowdisplaybreaks 

\begin{document}
\renewcommand{\thefootnote}{\fnsymbol{footnote}}
\newpage\normalsize
    \pagestyle{plain}
    \setlength{\baselineskip}{4ex}\par
    \setcounter{footnote}{0}
    \renewcommand{\thefootnote}{\arabic{footnote}}
\newcommand{\preprint}[1]{%
\begin{flushright}
\setlength{\baselineskip}{3ex} #1
\end{flushright}}
\renewcommand{\title}[1]{%
\begin{center}
    \LARGE #1
\end{center}\par}
\renewcommand{\author}[1]{%
\vspace{2ex}
{\Large
\begin{center}
    \setlength{\baselineskip}{3ex} #1 \par
\end{center}}}
\renewcommand{\thanks}[1]{\footnote{#1}}
\begin{flushright}
    \today
\end{flushright}
\vskip 0.5cm

\begin{center}
{\bf \Large Squark Decays in MSSM Under the Cosmological Bounds}
\end{center}
\vspace{1cm}
\begin{center}
Z. Z. Aydin
  \footnote{e-mail address: Z.Zekeriya.Aydin@eng.ankara.edu.tr }
  and
L. Selbuz \footnote{e-mail address: selbuz@eng.ankara.edu.tr }
    \end{center}
\vspace{1cm}
\begin{center}
 Department of Engineering Physics, Faculty of Engineering,
Ankara University, \\
06100  Tandogan-Ankara, Turkey \\
\end{center}
\vspace{1cm}
\begin{abstract}
We present the numerical investigation of the fermionic two-body
decays of squarks in the Minimal Supersymmetric Standard Model with
complex parameters. In the analysis we particularly take into
account the cosmological bounds imposed by WMAP data. We find that
the phase dependences of the decay widths of the third family
squarks, as well as those of the first and second families, are
quite significant which can provide viable probes of additional CP
sources. We plot the CP phase dependences for each fermionic
two-body decay channel of squarks $\tilde q_i$ (i=1,2 ,
q=u,d;c,s;t,b) and speculate about the branching ratios and total
(two-body) decay widths.
\end{abstract}


\section{Introduction}
\setcounter{equation}{0}
The experimental HEP frontier is soon reaching TeV energies and most
of physicists expect that just there theoretically proposed Higgs
bosons and super partners are waiting to be discovered. There are
many reasons to be so optimistic. First of all, in spite of its
remarkable successes, the Standard Model has to be extended into a
more complete theory which should solve the hierarchy problem and
stabilize the Higgs boson mass against radiative corrections. The
most attractive extension to realize these objectives is
supersymmetry (SUSY) \cite{haber}. Its minimal version (MSSM)
requires a non-standard Higgs sector \cite{higgssector} which
introduces additional sources of
CP-violation\cite{Dugan:1984qf,Masiero:2002xj} beyond the
$\delta_{CKM}$ phase \cite{Cabibbo:yz}. The plethora of CP-phases
also influences the decays and mixings of B mesons (as well as D and
K mesons). The present experiments at BABAR, Tevatron and KEK and
the one to start at the LHC will be able to measure various decay
channels to determine if there are supersymmetric sources of CP
violation. In particular, CP-asymmetry and decay rate of $B
\rightarrow X_s \gamma$ form a good testing ground for low-energy
supersymmetry with CP violation \cite{bmeson}. The above-mentioned
additional CP-phases explain the cosmological baryon asymmetry of
the universe and the lightest SUSY particle could be an excellent
candidate for cold dark matter in the universe
\cite{Goldberg:1983nd,Ellis:1983ew}.

In the case of exact supersymmetry, all scalar particles would have
to have same masses with their associated SM partners. Since none of
the superpartners has been discovered, supersymmetry must be broken.
But in order to preserve the hierarchy problem solved the
supersymmetry must be broken softly. This leads to a reasonable mass
splittings between known particles and their superpartners, i.e. to
the superpartners masses around 1 TeV.

The precision experiments by Wilkinson Microwave Anisotropy Probe
(WMAP) \cite{Wilkinson} have put the following constraint on the
relic density of cold dark matter \footnote{In our calculation, we
have used WMAP-allowed bands in the plane $ M_1-\varphi$ which are
based on 1st year data. Now the WMAP 3rd year data is also available
\cite{Spergel:2006hy}, but the new WMAP + SDSS combined value for
relic density of dark matter does not change the numerical results
in Ref. \cite{Belanger}, namely the WMAP-allowed bands. See "Note
added" section of Ref. \cite{Belanger}.}
\begin{equation}\label{mkl0}
0.0945 < \Omega_{CDM}h^2 < 0.1287
\end{equation}

Recently, in the light of this cosmological constraint an extensive
analysis of the neutralino relic density in the presence of SUSY-CP
phases has been given by B\'{e}langer \textit{et al.}
\cite{Belanger}.

In this study we present the numerical investigation of the
fermionic two-body decays of squarks in MSSM with complex SUSY
parameters. Actually, we had performed a short study in this
direction for the third family squarks \cite{selbuz} incorporating
all the existing bounds on the SUSY parameter space by utilizing the
study by Belanger et al. \cite{Belanger} . This investigation showed
us that the effects of $M_1$ and its phase $\varphi_{U(1)}$ on the
decay widths of $\tilde t_{1,2}$ and $\tilde b_{1,2}$ are quite
significant. Now we extend it to all three families. Although the
SUSY parameters $\mu, M_1, M_2$ and $A_f$ are in general complex, we
assume that $\mu, M_2$ and $A_f$ are real, but $M_1$ and its phase
$\varphi_{U(1)}$ take values on the WMAP-allowed bands. These bands
also satisfy the EDM bounds \cite{edms}. The experimental upper
limits on the EDMs of electron, neutron and the $^{299}Hg$ and the
$^{205}Tl$ atoms  may impose constraints on the size of the SUSY
CP-phases \cite{Ellis:1982tk,Barger:2001nu}. However, these
constraints are highly model dependent. This means that it is
possible to suppress the EDMs without requiring the various SUSY
CP-phases be small. For example, in the MSSM assuming strong
cancelations between different contributions \cite{Ibrahim:1997nc},
the phase of $\mu$ is restricted to $|\varphi_\mu|< \pi/10$, but
there is no such restriction on the phases of $M_1$ and $A_f$. In
addition, we evaluate the parameter $M_2$ via the relation $
M_2=(3/5)|M_1|(\tan\theta_W)^{-2}$ which can be derived by assuming
gaugino mass unification purely in the electroweak sector of MSSM.
It is very important to insert the WMAP-allowed band in the plane $
M_1-\varphi$ into the numerical calculations instead of taking one
fixed $M_1$ value for all $\varphi$-phases, because, for example, on
the allowed band for $\mu=200$ GeV, $M_1$ starts from 140 GeV for
$\varphi=0$ and increasing monotonously it becomes 165 GeV for
$\varphi=\pi$. In Ref.\cite{Belanger} two WMAP-allowed band plots
are given, one for $\mu=200$ GeV and the other for $\mu=350$ GeV.
For both plots the other parameters are fixed to be $\tan \beta=10$,
$ m_{H^+}=1$ TeV, $A_f=1.2 $ TeV, $\varphi_\mu$=$\varphi_{A_f}$=0.
We here simply choose generic masses for squarks as $m_{\tilde
q_2}$=1000 GeV and $m_{\tilde q_1}$=750 GeV (q=u,d,c,s,t,b),
although we had chosen more reasonable mass values in our previous
note \cite{selbuz}. Taking different sets of values for squark
masses around 1 TeV does not affect much neither the phase
dependences nor the decay widths. As a second possibility, we take
masses of 10 TeV for the first and second family squarks.


\section{Squark Masses, Mixing and Two-body Decay Widths }
\setcounter{equation}{0}

\subsection{Masses and mixing in squark sector}
The superpartners of the SM fermions with left and right helicity
are the left and right sfermions. In the case of top squark (stop)
and bottom squark (sbottom) the left and right states are in general
mixed. Therefore, the sfermion mass terms of the Lagrangian are
described in the basis ($\tilde q_{L}$,$\tilde q_{R}$) as
\cite{Ellis:1983ed,Gunion:1984yn}
\begin{equation}\label{mkl1}
 {\cal L}_M^{\tilde q }= -({\tilde q}_L^{\dag}{\tilde q}_R^{\dag})\left(
 \begin{array}{cc}
 M_{L L}^{2}& M_{L R}^{2}\\[1.ex]
 M_{R L}^{2} & M_{R R}^{2}
 \end{array}
 \right)
 \left(
\begin{array}{c}
\tilde q_L\\ [1.ex] \tilde q_R
\end{array}
\right)
\end{equation}
with
\begin{eqnarray}\label{mkl2}
M_{L L}^{2}&=&M_{\tilde Q}^{2}+(I_{3L}^{q}-e_q\sin^2\theta_W)\cos(2
\beta)m_{z}^{2}+m_{q}^{2}\\
M_{R R}^{2}&=&M_{\tilde Q'}^{2}+e_q\sin^2\theta_W\cos(2
\beta)m_{z}^{2}+m_{q}^{2}\\\label{mkl3} M_{R L}^{2}&=&(M_{L
R}^{2})^{*}=m_q(A_q-\mu^{*}(\tan\beta)^{-2I_{3L}^{q}})\label{mkl4}
\end{eqnarray}
where $m_q$, $e_q$, $I_{3L}^{q}$ and $\theta_W$ are the mass,
electric charge, weak isospin of the quarks and the weak mixing
angle, respectively. $\tan\beta=v_2/v_1$ with $v_i$ being the vacuum
expectation values of the Higgs fields $H_i^{0}$, $ i=1,2$. The
soft-breaking parameters $M_{\tilde Q}$, $M_{\tilde Q'}=M_{\tilde U}
(M_{\tilde D})$ for up components (down components), $A_q$ involved
in Eqs. (2.2-2.4) can be evaluated for our numerical calculations
using the following relations
\begin{eqnarray}\label{mkl5}
M_{\tilde Q}^{2}&=&\frac{1}{2}{\left(m_{\tilde q_1}^{2}+m_{\tilde
q_2}^{2} \pm\sqrt{(m_{\tilde q_2}^{2}-m_{\tilde q_1}^{2})^2-4m_q^{2}
|A_q-\mu^{*}\cot\beta|^2}\right)}\nonumber \\
&&-(\frac{1}{2}-\frac{2}{3}\sin^2\theta_W)\cos(2\beta)m_{z}^{2}-m_{q}^{2}\\
\label{mkl6} M_{\tilde U}^{2}&=&\frac{1}{2}{\left(m_{\tilde
q_1}^{2}+m_{\tilde q_2}^{2} \mp\sqrt{(m_{\tilde q_2}^{2}-m_{\tilde
q_1}^{2})^2-4m_q^{2}
|A_q-\mu^{*}\cot\beta|^2}\right)}\nonumber \\
&&-\frac{2}{3}\sin^2\theta_W\cos(2\beta)m_{z}^{2}-m_{q}^{2}\\
\label{mkl8} M_{\tilde D}^{2}&=&\frac{1}{2}{\left(m_{\tilde
q_1}^{2}+m_{\tilde q_2}^{2} \mp\sqrt{(m_{\tilde q_2}^{2}-m_{\tilde
q_1}^{2})^2-4m_q^{2}
|A_q-\mu^{*}\cot\beta|^2}\right)}\nonumber \\
&&+\frac{1}{3}\sin^2\theta_W\cos(2\beta)m_{z}^{2}-m_{q}^{2}
\end{eqnarray}

The squark mass eigenstates $\tilde q_1$ and $\tilde q_2$
can be obtained from the weak states $\tilde q_L$ and $\tilde q_R$ via the $\tilde q$-mixing matrix
\begin{equation}\label{mkl10}
 {\cal R}^{\tilde q }=\left(
 \begin{array}{cc}
 e^{i\varphi_{\tilde q}}\cos\theta_{\tilde q}& \sin\theta_{\tilde q}\\[1.ex]
 -\sin\theta_{\tilde q} & e^{-i\varphi_{\tilde q}}\cos\theta_{\tilde q}
 \end{array}
 \right)
\end{equation}
where
 \begin{equation}\label{mkl9}
 \varphi_{\tilde q}=\arg[M_{R
 L}^{2}]=\arg[A_q-\mu^{*}(\tan\beta)^{-2I_{3L}^{q}}]
 \end{equation}
 and
\begin{equation}\label{mkl11}
 \cos\theta_{\tilde q}=\frac{-|M_{L R}^{2}|}
 {\sqrt{|M_{L R}^{2}|^2+
 (m_{\tilde q_1}^{2}-M_{L L}^{2})^2}}, \qquad
\sin\theta_{\tilde q}=\frac{M_{L L}^{2}-m_{\tilde q_1}^{2}}
 {\sqrt{|M_{L R}^{2}|^2+
 (m_{\tilde q_1}^{2}-M_{L L}^{2})^2}}
\end{equation}
One can easily get the following squark mass eigenvalues by diagonalizing the mass matrix in Eq. (2.1):
\begin{equation}\label{mkl12}
 m_{\tilde q_{1,2}}^{2}=\frac{1}{2}
{\left(M_{L L}^{2}+M_{R R}^{2} \mp\sqrt{(M_{L L}^{2}-M_{R
R}^{2})^2+4|M_{L R }^{2}|^2} \right)} ,\qquad m_{\tilde q_1}<
m_{\tilde q_2}
\end{equation}

Note that the left-right mixing is not significant for the first and
second generations; i.e. , $\tilde q_1\simeq\tilde q_R$ and $\tilde
q_2\simeq-\tilde q_L$ (q=u,d,c,s).

We might add a comment about the possibility of a flavor mixing, for
example, between the second and third squark families. In this case,
the sfermion mass matrix in Eq. (2.1) becomes a 4x4 matrix in the
basis ($\tilde c_L$, $\tilde c_R$, $\tilde t_L$, $\tilde t_R$). Then
one obtains squark mass eigenstates ($\tilde c_1$, $\tilde c_2$,
 $\tilde t_1$, $\tilde t_2$) from these weak states, and analyzes
their decays by utilizing procedures similar to the ones indicated
in the text. The problem with flavor violation effects is that their
inclusion necessarily correlates B, D and K physics with direct
sparticle searches at colliders. Moreover, it has been shown that,
with sizeable supersymmetric flavor violation, even the Higgs
phenomenology at the LHC correlates with that of the rare processes
\cite{flavor}. In this work we have neglected such effects; however,
we emphasize that inclusion of such effects can give important
information on mechanism that breaks supersymmetry via decay
products of squarks.

\subsection{Fermionic decay widths of squarks  }
The quark-squark-chargino and quark-squark-neutralino Lagrangians
have been first given in Ref. \cite{haber}. Here we use them in
notations of Ref. \cite{Bartl}:
\begin{eqnarray}\label{mkl13}
{\cal L}_{q \tilde q \tilde \chi^{+}}=g\bar{u}(\ell_{i j}^{\tilde
d}P_R + k_{i j}^{\tilde d}P_L){\tilde \chi}_j^{+}{\tilde
d}_i+g\bar{d}(\ell_{i j}^{\tilde u}P_R + k_{i j}^{\tilde
u}P_L){\tilde \chi}_j^{+c}{\tilde u}_i+h.c.
\end{eqnarray}
and
\begin{eqnarray}\label{mkl18}
{\cal L}_{q \tilde q \tilde \chi^{0}} =g\bar{q}(a_{i k}^{\tilde
q}P_R + b_{i k}^{\tilde q}P_L){\tilde \chi}_k^{0}{\tilde q}_i+h.c.
\end{eqnarray}
where u ($\tilde u$) stands for up-type (s)quark and d ($\tilde d$)
stands for down-type s(quark). We also borrow the formulas for the
partial decay widths of $\tilde q_i$ ( $\tilde q_i$=$\tilde
t_i$,$\tilde u_i$,$\tilde b_i$ and $\tilde d_i$) into quark-chargino
(or neutralino) from Ref. \cite{Bartl}:

\begin{eqnarray}\label{mkl22}
\Gamma(\tilde q_i\rightarrow q^{'}+\tilde
\chi_k^\pm)&=&\frac{g^2\lambda^{1/2}( m_{\tilde
q_{i}}^2,m_{q'}^{2},m_{\tilde \chi_k^\pm}^2)}{16\pi m_{\tilde
q_{i}}^3}\times \nonumber\\
&& {\left[{\left(|k_{i k}^{\tilde q}|^2+|\ell_{i k}^{\tilde
q}|^2\right)}(m_{\tilde q_{i}}^2-m_{q'}^{2}-m_{\tilde
\chi_k^\pm}^2)-4Re(k_{i k}^{\tilde q *}\ell_{i k}^{\tilde
q})m_{q'}m_{\tilde \chi_k^\pm}\right]}
\end{eqnarray}
and
\begin{eqnarray}\label{mkl23}
\Gamma(\tilde q_i\rightarrow q+\tilde
\chi_k^0)&=&\frac{g^2\lambda^{1/2}( m_{\tilde
q_{i}}^2,m_{q}^{2},m_{\tilde \chi_k^0}^2)}{16\pi m_{\tilde
q_{i}}^3}\times \nonumber\\
&& {\left[{\left(|a_{i k}^{\tilde q}|^2+|b_{i k}^{\tilde
q}|^2\right)}( m_{\tilde q_{i}}^2-m_{q}^{2}-m_{\tilde
\chi_k^0}^2)-4Re(a_{i k}^{\tilde q *}b_{i k}^{\tilde
q})m_{q}m_{\tilde \chi_k^0}\right]}
\end{eqnarray}
with $\lambda(x,y,z)=x^{2}+y^{2}+z^{2}-2(xy+xz+yz)$.

The explicit forms of $\ell_{i k}^{\tilde q}$, $k_{i k}^{\tilde q}$
are

\begin{equation}\label{mkl15}
\ell_{i k}^{\tilde q}=-{\cal R}_{i1}^{{\tilde q}^{*}}V_{k1}+Y_U{\cal
R}_{i2}^{{\tilde q}^{*}}V_{k2},   \qquad   k_{i k}^{\tilde q}={\cal
R}_{i1}^{{\tilde q}^{*}}Y_DU_{k2}^{*},
\end{equation}
for up squarks $\tilde u$, $\tilde c$, $\tilde t$ and
\begin{equation}\label{mkl16}
\ell_{i k}^{\tilde q}=-{\cal R}_{i1}^{{\tilde q}^{*}}U_{k1}+Y_D{\cal
R}_{i2}^{{\tilde q}^{*}}U_{k2},   \qquad   k_{i k}^{\tilde q}={\cal
R}_{i1}^{{\tilde q}^{*}}Y_UV_{k2}^{*},
\end{equation}
for down squarks $\tilde d$, $\tilde s$, $\tilde b$ (U=$u$, $c$, $t$
; D=$d$, $s$, $b$). And $a_{i k}^{\tilde q}$, $b_{i k}^{\tilde q}$
are given as
\begin{equation}\label{mkl19}
a_{i k}^{\tilde q}=\sum_{n=1}^{2}({\cal R}_{in}^{{\tilde
q}})^{*}{\cal A}_{kn}^{q},    \qquad   b_{i k}^{\tilde
q}=\sum_{n=1}^{2}({\cal R}_{in}^{{\tilde q}})^{*}{\cal B}_{kn}^{q},
 \qquad
\end{equation}
where
\begin{equation}
{\cal A}_k^{q}=
\begin{pmatrix}
\ f_{Lk}^q \\ \ h_{Rk}^q\\
\end{pmatrix}, \qquad
{\cal B}_k^{q}=
\begin{pmatrix}
\ h_{Lk}^q \\ \ f_{Rk}^q\\
\end{pmatrix}
\end{equation}
Here, the components for up squarks $\tilde u$, $\tilde c$, $\tilde
t$ are
\begin{eqnarray}\label{mkl20}
f_{Lk}^{q}&=&-\frac{1}{\sqrt{2}}(N_{k2}+\frac{1}{3}\tan\theta_WN_{k1})\nonumber \\
f_{Rk}^{q}&=&\frac{2\sqrt{2}}{3}\tan\theta_WN_{k1}^{*}\nonumber \\
h_{Lk}^{q}&=&(h_{Rk}^{q})^{*}=-Y_UN_{k4}^{*}
\end{eqnarray}
 and for down squarks $\tilde d$, $\tilde s$, $\tilde b$
\begin{eqnarray}\label{mkl21}
f_{Lk}^{q}&=&\frac{1}{\sqrt{2}}(N_{k2}-\frac{1}{3}\tan\theta_WN_{k1})\nonumber \\
f_{Rk}^{q}&=&-\frac{\sqrt{2}}{3}\tan\theta_WN_{k1}^{*}\nonumber \\
h_{Lk}^{q}&=&(h_{Rk}^{q})^{*}=-Y_DN_{k3}^{*}
\end{eqnarray}

We have to point out that although at the loop level the SUSY-QCD
corrections could be important, our analysis here are merely at tree
level, as can be seen from Eqs. (2.13) and (2.14). In this work we
content with tree-level amplitudes as we aim at determining the
phase-sensitivities of the decay rates, mainly.

\section{Two-body decays of first and second generations squarks }

Now we present the dependences of the $\tilde u_1$, $\tilde u_2$,
$\tilde d_1$ and $\tilde d_2$ two-body decay widths on the phase
$\varphi_{U(1)}$ for $ \mu =200 $ GeV and for $ \mu =350 $ GeV. Here
we choose the values for the masses  ($m_{\tilde u_2}=m_{\tilde
d_2}$, $m_{\tilde u_1}=m_{\tilde d_1}$ $m_{\tilde \chi_1^\pm}$,
$m_{\tilde \chi_2^\pm}$, $m_{\tilde \chi_1^0}$) = (1000 GeV, 750
GeV, 180 GeV, 336 GeV, 150 GeV) for  $ \mu =200 $ GeV and
($m_{\tilde u_2}=m_{\tilde d_2}$, $m_{\tilde u_1}=m_{\tilde d_1}$
$m_{\tilde \chi_1^\pm}$, $m_{\tilde \chi_2^\pm}$, $m_{\tilde
\chi_1^0}$) = (1000 GeV, 750 GeV, 340 GeV, 680 GeV, 290 GeV) for $
\mu =350 $ GeV. Note that although the neutralino and chargino
masses vary with $\varphi_{U(1)}$, these variations are not large.
Therefore, as a final state particle (i.e., on mass-shell), we have
chosen fixed (average) mass values for charginos and neutralinos.
Fig.1(a) and Fig.1(b) show the partial decay widths of the channels
$ \tilde u_2\rightarrow d\tilde \chi_1^+
 $, $ \tilde u_2\rightarrow d\tilde \chi_2^+
 $, $ \tilde u_2\rightarrow u\tilde \chi_1^0
 $, $ \tilde u_1\rightarrow u\tilde \chi_1^0
 $ and  $ \tilde d_2\rightarrow u\tilde \chi_1^-
 $, $ \tilde d_2\rightarrow u\tilde \chi_2^-
 $, $ \tilde d_2\rightarrow d\tilde \chi_1^0$,
$ \tilde d_1\rightarrow d\tilde \chi_1^0$, respectively for $ \mu
=200 $ GeV. [ We plot the same processes for $ \mu =350 $ GeV in
Fig.2(a) and Fig.2(b) ]. The decay channels $ \tilde u_1\rightarrow
d\tilde \chi_1^+
 $ and $ \tilde u_1\rightarrow d\tilde \chi_2^+
 $ are absent because the mixing elements (${\cal R}_{ij }$)
 and the Yukawa couplings $Y_u$, $Y_d$ in $\ell_{i j}^{\tilde u}$
and  $k_{i j}^{\tilde u}$ parameters are zero, whereas the channel
$ \tilde u_1\rightarrow u\tilde \chi_1^0
 $ is present, because the parameter $b_{1 1}$ is not zero.
 Because of a similar reason the phase dependence of  $ \tilde u_1\rightarrow
u\tilde \chi_1^0
 $ channel is very little.

 The branching ratios for $ \tilde u_2$ are roughly $B(\tilde u_2\rightarrow d\tilde \chi_2^+)$ :
 $B(\tilde u_2\rightarrow d\tilde\chi_1^+)$ :
$B(\tilde u_2\rightarrow u\tilde\chi_1^0)$ $\approx$ 5 : 2 : 0.05.
for the case $ \mu =200 $ GeV. From Fig.1(b) one can see that
$B(\tilde d_2\rightarrow u\tilde \chi_2^-)$ : $B(\tilde
d_2\rightarrow u\tilde\chi_1^-)$ : $B(\tilde d_2\rightarrow
d\tilde\chi_1^0)$ $\approx$ 6 : 1 : 0.2 for the case $ \mu =200 $
GeV. The width  $ \Gamma( \tilde u_2\rightarrow d\tilde \chi_2^+ )$
increases as the phase increases from 0 to $\pi$, but $ \Gamma(
\tilde u_2\rightarrow d\tilde \chi_1^+ )$ decreases as
$\varphi_{U(1)}$ increases, both showing significant dependence on
the phase. The phase dependence is more significant for the decay
channel $ \tilde u_2\rightarrow u\tilde \chi_1^0$ (its decay width
increases from 0.0006 GeV to 0.15 GeV as $\varphi_{U(1)}$ increases
from 0 to $\pi$) but its branching ratio is too small to observe it.

The channels $\tilde d_1\rightarrow u\tilde \chi_1^-$ and $\tilde
d_1\rightarrow u\tilde \chi_2^-$ are absent because the related
mixing elements and Yukawa couplings are zero ( therefore $\ell_{1
1}^{\tilde d}$, $k_{1 1}^{\tilde d}$,  $\ell_{1 2}^{\tilde d}$,
$k_{1 2}^{\tilde d}$ become zero); whereas the channel $\tilde
d_1\rightarrow d\tilde \chi_1^0$ is present (since  $b_{1 1}^{\tilde
d}$$\neq$0), but it is little with respect to $\tilde d_2$ decay
channels. Its phase dependence is also not so strong ( from
$\varphi_{U(1)}$=0 to $\pi$, its width increases only from 0.264 GeV
to 0.271 GeV ). Therefore, the $\tilde d_1$ decay can be ignored in
comparison with the $\tilde d_2$ decay.

The decay pattern of the second family squarks coincides with that
of the first sfamily, because we can set $m_s$$\approx$0  and
$m_c$$\approx$0 like  $m_u$$\approx$0 and $m_d$$\approx$0 in
comparison with the masses $m_{\tilde q_2}$=1000 GeV, $m_{\tilde
q_1}$=750 GeV ( The use of $m_s$$\approx$0.1 GeV and
$m_c$$\approx$1.5 GeV does change nothing).

In the analysis of cosmological constraints by Belanger et al. they
take masses of 10 TeV for the first and second generations of
squarks. As an academic exercise, we have repeated the above
calculations only changing the first and second squarks masses to
$m_{\tilde u_2}=m_{\tilde d_2}$=10 TeV and $m_{\tilde u_1}=m_{\tilde
d_1}$=9.5 TeV. We obtained again similar phase dependences, but
larger ($\sim$ one order) width values because of huge phase spaces.
\section{Two-body decays of third generation squarks  }

In this section we plot the dependences of the $\tilde t_1$, $\tilde
t_2$, $\tilde b_1$ and $\tilde b_2$ partial decay widths on $
\varphi_{U(1)}$ for $ \mu =200 $ GeV and for $ \mu =350 $ GeV. Here
we simply choose mass values for the third family squarks as
$m_{\tilde t_2}$=$m_{\tilde b_2}$=1000 GeV and $m_{\tilde
t_1}$=$m_{\tilde b_1}$=750 GeV in order to compare the results with
those of the first two families. We had taken more reasonable mass
spectrum for squarks in our previous study \cite {selbuz}. The other
masses are ($m_{\tilde \chi_1^\pm}$, $m_{\tilde \chi_2^\pm}$,
$m_{\tilde \chi_1^0}$) = (180 GeV, 336 GeV, 150 GeV) for  $ \mu =200
$ GeV and ($m_{\tilde \chi_1^\pm}$, $m_{\tilde \chi_2^\pm}$,
$m_{\tilde \chi_1^0}$) = (340 GeV, 680 GeV, 290 GeV) for $ \mu =350
$ GeV.

 For both sets of values by calculating the  $M_{\tilde
Q}$ and $M_{\tilde U}$ values corresponding to $m_{\tilde t_1}$ and
$m_{\tilde t_2}$, we plot the decay widths  for $M_{\tilde Q} \geq
M_{\tilde U}$  and $M_{\tilde Q} < M_{\tilde U}$, separately.
Fig.3(a) and Fig.3(b) show the partial decay widths of the channels
$ \tilde t_1\rightarrow b\tilde \chi_1^+
 $,  $ \tilde t_1\rightarrow b\tilde \chi_2^+ $   $ \tilde t_1\rightarrow t\tilde \chi_1^0 $,
 $ \tilde t_2\rightarrow b\tilde \chi_1^+
 $,   $ \tilde t_2\rightarrow b\tilde \chi_2^+ $  and  $ \tilde t_2\rightarrow t\tilde \chi_1^0
 $   as a function of $ \varphi_{U(1)}$ for $ \mu =200$ GeV assuming $M_{\tilde Q} > M_{\tilde U}$ and
 $M_{\tilde Q} < M_{\tilde U}$, respectively. In these figures
 significant dependences on $ \varphi_{U(1)}$  phase are seen. $ \Gamma( \tilde t_1\rightarrow t\tilde \chi_1^0 )$ and
$\Gamma(\tilde t_1\rightarrow b\tilde
 \chi_1^+)$ decay widths increase as $ \varphi_{U(1)}$ increases
 from 0 to $\pi$, but $\Gamma(\tilde t_1\rightarrow b\tilde
 \chi_2^+)$ width decreases as  $\varphi_{U(1)}$ increases. On the
 other hand, $ \Gamma( \tilde t_2\rightarrow t\tilde \chi_1^0 )$
 decrease for both  $M_{\tilde Q} > M_{\tilde U}$ and  $M_{\tilde Q} < M_{\tilde
 U}$ cases as  $\varphi_{U(1)}$ increases; $\Gamma(\tilde t_2\rightarrow b\tilde
 \chi_1^+)$ decreases for $M_{\tilde Q} > M_{\tilde U}$ but
 increases for $M_{\tilde Q} < M_{\tilde U}$ and $\Gamma(\tilde t_2\rightarrow b\tilde
 \chi_2^+)$ increases for $M_{\tilde Q} > M_{\tilde U}$ but
 decreases for $M_{\tilde Q} < M_{\tilde U}$.

 The branching ratios for $\tilde t_2$ are roughly $B(\tilde t_2\rightarrow b\tilde
 \chi_1^+)$ : $B(\tilde t_2\rightarrow t\tilde
 \chi_1^0)$ : $B(\tilde t_2\rightarrow b\tilde
 \chi_2^+)$ $\approx$ 12 : 2.5 : 1. This simply reflects both the large phase space
 and large Yukawa coupling for the decay $\tilde t_2\rightarrow b\tilde
 \chi_1^+$.

 In Figs. 4(a) and 4(b) you see the same partial decay widths for  $ \mu =350 $
 GeV. They, too, show the significant
 dependences on CP-violation phase. For both $M_{\tilde Q} > M_{\tilde U}$ and  $M_{\tilde Q} < M_{\tilde
 U}$ cases; $ \Gamma( \tilde t_1\rightarrow t\tilde \chi_1^0 )$ and
$\Gamma(\tilde t_1\rightarrow b\tilde
 \chi_1^+)$ decay widths increase as $ \varphi_{U(1)}$ increases
 from 0 to $\pi$, but $\Gamma(\tilde t_1\rightarrow b\tilde
 \chi_2^+)$ width decreases as  $\varphi_{U(1)}$ increases. Besides
 that $ \Gamma( \tilde t_2\rightarrow t\tilde \chi_1^0 )$ and $\Gamma(\tilde t_2\rightarrow b\tilde
 \chi_1^+)$ widths decrease but $\Gamma(\tilde t_2\rightarrow b\tilde
 \chi_2^+)$ width increases as $ \varphi_{U(1)}$ increases
 from 0 to $\pi$ for both $M_{\tilde Q} > M_{\tilde U}$ and  $M_{\tilde Q} < M_{\tilde
 U}$ cases. The branching ratios for $\tilde t_2$ are roughly $B(\tilde t_2\rightarrow b\tilde
 \chi_1^+)$ : $B(\tilde t_2\rightarrow t\tilde
 \chi_1^0)$ : $B(\tilde t_2\rightarrow b\tilde
 \chi_2^+)$ $\approx$ 8 : 2 : 1.

 For  $ \mu =350 $ GeV the WMAP-allowed band \cite{Belanger} takes place in larger $M_1$
 values ($ \sim 305-325$ GeV) leading to larger
 chargino and neutralino masses. This naturally leads very small
 decay width for $\tilde t_1\rightarrow b\tilde
 \chi_2^+$. The decay width of the process $\tilde t_2\rightarrow
 b\tilde \chi_1^+$ is the largest one among the $\tilde t_2$
 channels and the branching ratios are $B(\tilde t_2\rightarrow b\tilde
\chi_1^+)$ : $B(\tilde t_2\rightarrow t\tilde \chi_1^0)$ : $B(\tilde
t_2\rightarrow b\tilde \chi_2^+)$ $\approx$ 8 : 3 : 1. The decay
$\tilde t_2\rightarrow t\tilde\chi_1^{0}$ shows strong phase
dependence.

We give sbottom decay widths as a function of  $ \varphi_{U(1)}$ in
Figs. 5(a)-(b)(for $ \mu =200 $ GeV ) and in Figs. 6(a)-(b)(for $
\mu =350 $ GeV ). $\Gamma(\tilde b_2\rightarrow b\tilde
 \chi_1^0)$ is smaller than $\Gamma(\tilde b_2\rightarrow t\tilde
 \chi_i^-)$ in spite of large phase space, because in $ \tilde b_2\rightarrow b\tilde \chi_1^0
 $ decay only $Y_b$ coupling enters which is very small in comparison with $Y_t$. The dependences of the phase $
 \varphi_{U(1)}$ in sbottom decays are similar to those in stop
 decays.

 For the case $M_{\tilde Q} > M_{\tilde D}$,  $\tilde b_2$ decay is
 7-10 times larger than that of  $\tilde b_1$, whereas for $M_{\tilde Q} < M_{\tilde
 D}$ the reverse is true. The branching ratios for $\tilde b_2$ decays are  $B(\tilde b_2\rightarrow t\tilde \chi_1^-)$ :
 $B(\tilde b_2\rightarrow t\tilde\chi_2^-)$ : $B(\tilde b_2\rightarrow
 b\tilde\chi_1^0)$  $\approx$ 14 : 10 : 0.5 for $ \mu =200 $ GeV and $M_{\tilde Q} > M_{\tilde
 D}$, and similarly 14 : 3 : 0.5 for $ \mu =350 $ GeV and $M_{\tilde Q} > M_{\tilde
 D}$. While the process $\tilde b_2\rightarrow
 b\tilde\chi_1^0$ is suppressed more than one order its dependence
 on $\varphi_{U(1)}$ is prominent such that the value of decay
 width at  $\varphi_{U(1)}=0$ is 2 times larger than that at
 $\varphi_{U(1)}=\pi$.  $\varphi$-dependences of the processes $\tilde b_2\rightarrow t\tilde
 \chi_1^-$ and $\tilde b_2\rightarrow t\tilde \chi_2^-$ can be
 seen easily in Fig. 5(a).

The branching ratios for $\tilde b_1$ decays are  $B(\tilde
b_1\rightarrow t\tilde \chi_1^-)$ : $B(\tilde b_1\rightarrow
t\tilde\chi_2^-)$ : $B(\tilde b_1\rightarrow b\tilde\chi_1^0)$
$\approx$ 9 : 3 : 0.5 for $ \mu =200 $ GeV and  $M_{\tilde Q} <
M_{\tilde D}$ and $B(\tilde b_1\rightarrow t\tilde \chi_1^-)$ :
$B(\tilde b_1\rightarrow b\tilde\chi_1^0)$  $\approx$ 7 : 0.3 for $
\mu =350 $ GeV and $M_{\tilde Q} < M_{\tilde D}$.

\section{Discussion  }
\setcounter{equation}{0}

In this paper, we present the numerical investigation of the
fermionic two-body decays of squarks in three families in the
Minimal Supersymmetric Standard Model with complex parameters taking
into account the cosmological bounds imposed by WMAP data. Numerical
calculations of decay widths at tree level show significant
dependence on the CP phase $ \varphi_{U(1)}$ for the third family
squarks, as well as for the first and second families.

We have assumed the same mass of 750 GeV, for all squarks of type 1
and the same mass of 1000 GeV for all squarks of type 2. The decay
width values of the third family squarks and those of the first two
families differ only two to three times in favor of the third
family.

In the case of $ \mu =200 $ GeV all the channels of $\tilde u_2$
decay ($\tilde d_2$ decay), i.e. , $\tilde u_2\rightarrow d\tilde
 \chi_2^+$, $\tilde u_2\rightarrow d\tilde
 \chi_1^+$ and $\tilde u_2\rightarrow u\tilde
 \chi_1^0$  ($\tilde d_2\rightarrow u\tilde
 \chi_2^-$, $\tilde d_2\rightarrow u\tilde
 \chi_1^-$ and $\tilde d_2\rightarrow d\tilde
 \chi_1^0$)  are present and have very significant phase dependences,
 but the channel $\tilde u_2\rightarrow u\tilde
 \chi_1^0$ ($\tilde d_2\rightarrow d\tilde
 \chi_1^0$) has a very small branching ratio. At the $\tilde u_1$
 decay ($\tilde d_1$ decay) only one channel, i.e.,  $\tilde u_1\rightarrow u\tilde
 \chi_1^0$ ($\tilde d_1\rightarrow d\tilde
 \chi_1^0$) is open with a small branching ratio. The decay channels
 $\tilde u_1\rightarrow d\tilde
 \chi_1^+$ and $\tilde u_1\rightarrow d\tilde
 \chi_2^+$ ($\tilde d_1\rightarrow u\tilde
 \chi_1^-$ and $\tilde d_1\rightarrow u\tilde
 \chi_2^-$) are absent because of the reason explained in section 3.

 Very roughly speaking, for $ \mu =200 $ GeV and $M_{\tilde Q} > M_{\tilde
 U}$ or $M_{\tilde D}$ we find the following total (two-body) widths:
\begin{eqnarray}\label{mkl100}
\Gamma_{\text{total}}(\tilde u_2\,\text{or}\, \tilde c_2)&\approx&
\Gamma_{\text{total}}(\tilde d_2\,\text{or}\, \tilde s_2)\approx
7\,\text{GeV}
\\\Gamma_{\text{total}}(\tilde u_1 \,\text{or}\, \tilde c_1)&\approx& 1\,\text{GeV}
\\\label{mkl101}
\Gamma_{\text{total}}(\tilde d_1 \,\text{or}\,\tilde
s_1)&\approx&0.5\,\text{GeV}
\\\label{mkl102}
\Gamma_{\text{total} }(\tilde t_2)&\approx&15\,\text{GeV}
\\\label{mkl103}
\Gamma_{\text{total}}(\tilde b_2)&\approx& 24 \,\text{GeV}
\\\label{mkl104}
\Gamma_{\text{total}}(\tilde t_1)&\approx& 10 \,\text{GeV}
\\\label{mkl104}
\Gamma_{\text{total} }(\tilde b_1)&\approx& 3 \,\text{GeV}
\end{eqnarray}

From these results we see that although the decays of the third
family squarks are more important, the decays of the first two
families are not ignorable at all. For example, if we assume that
the probability of producing every kind of squarks in a
proton-proton collision (LHC) are more or less equal, then the ratio
of total decay width of the third family squarks to that of the
first two families would be approximately 2.

In the case of the choice of squarks masses of 10 TeV for the first
two families and 1 TeV for the third family which we call it as "an
academic exercise" (because there is no hope to produce the first
two family squarks in LHC) this ratio even reverses in favor of the
first two sfamilies.

\newpage
\begin{figure}
\includegraphics{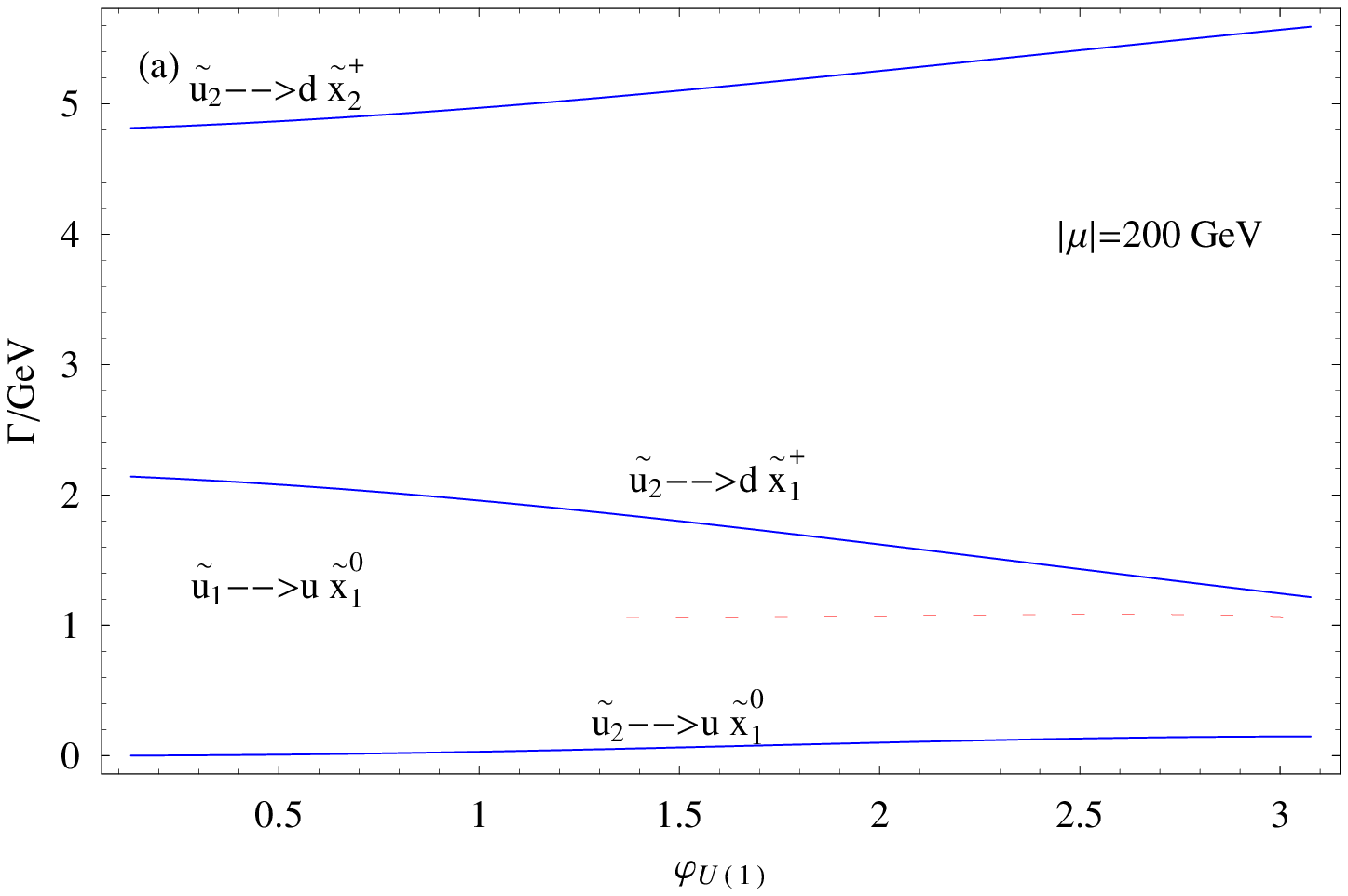} 
\label{figur1}
\end{figure}
\begin{figure}
\includegraphics{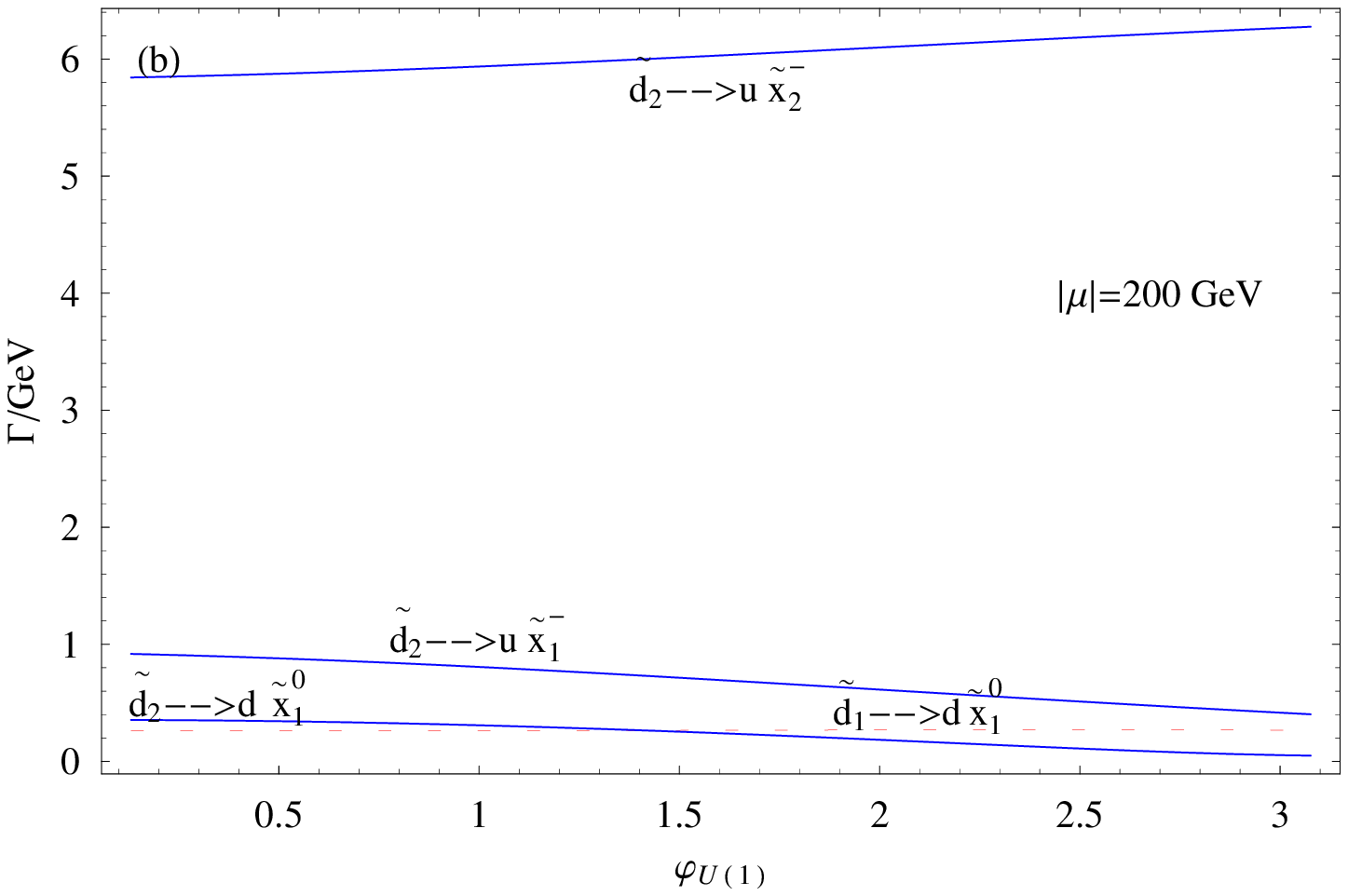} 
\caption{(a)-(b) Partial decay widths $\Gamma$ of the  $\tilde
u_{1,2}$ and $\tilde d_{1,2}$ decays for $\mu =200 $ GeV ,
$\tan\beta=10$, $A_u$=$A_d$=1.2  TeV,
$\varphi_\mu$=$\varphi_{A_u}$=$\varphi_{A_d}$=0,
  $m_{\tilde u_1}$=$m_{\tilde d_1}$$=750$ GeV and $m_{\tilde u_2}$=$m_{\tilde d_2}$$=1000$ GeV. } \label{figur2}
\end{figure}
\begin{figure}
\includegraphics{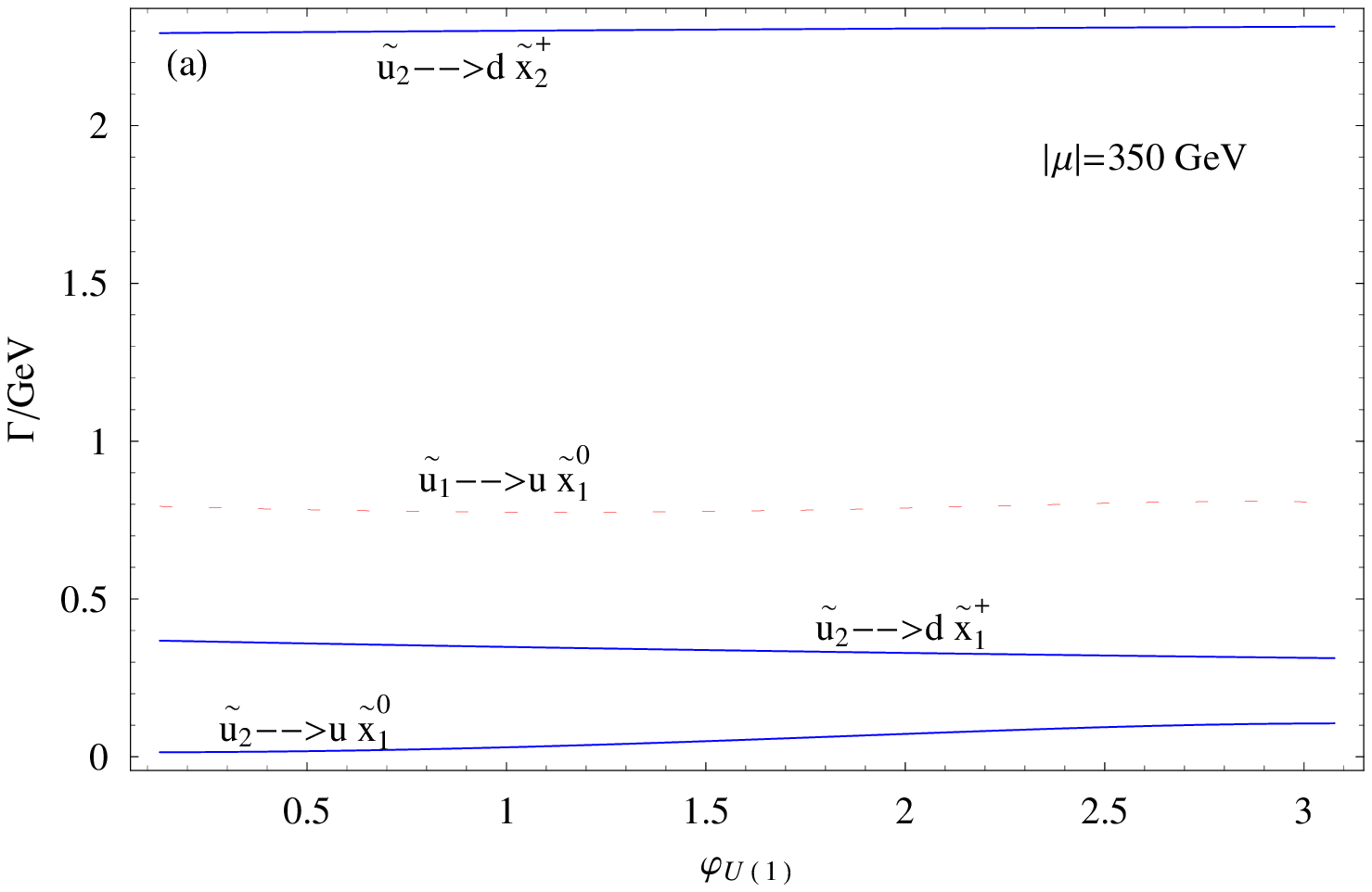} 
\label{figur3}
\end{figure}
\begin{figure}
\includegraphics{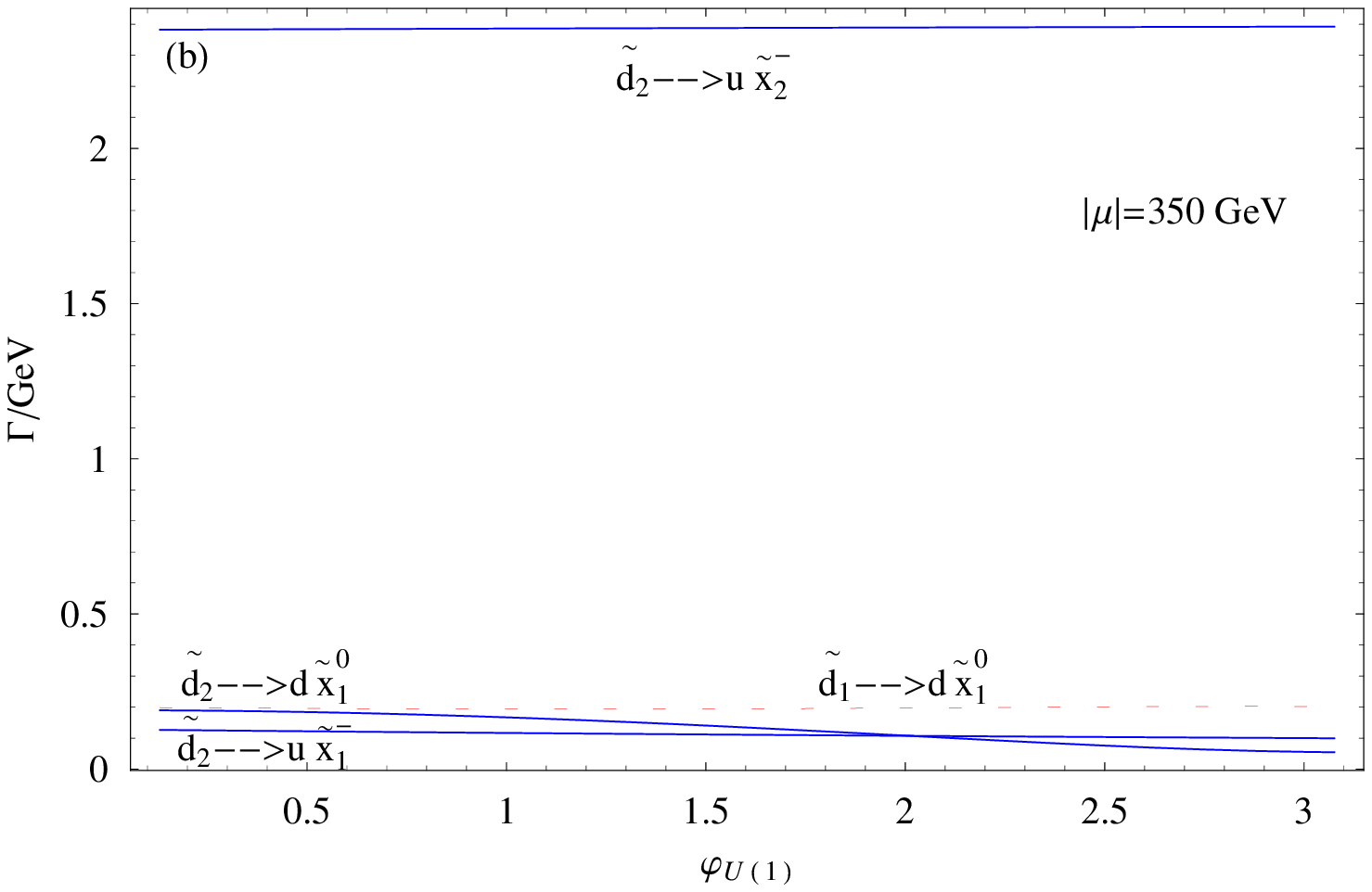} 
\caption{(a)-(b) Partial decay widths $\Gamma$ of the  $\tilde
u_{1,2}$ and $\tilde d_{1,2}$ decays for $\mu =350 $ GeV ,
$\tan\beta=10$, $A_u$=$A_d$=1.2  TeV,
$\varphi_\mu$=$\varphi_{A_u}$=$\varphi_{A_d}$=0,
  $m_{\tilde u_1}$=$m_{\tilde d_1}$$=750$ GeV and $m_{\tilde u_2}$=$m_{\tilde d_2}$$=1000$ GeV. }
\label{figur4}
\end{figure}
\begin{figure}
\includegraphics{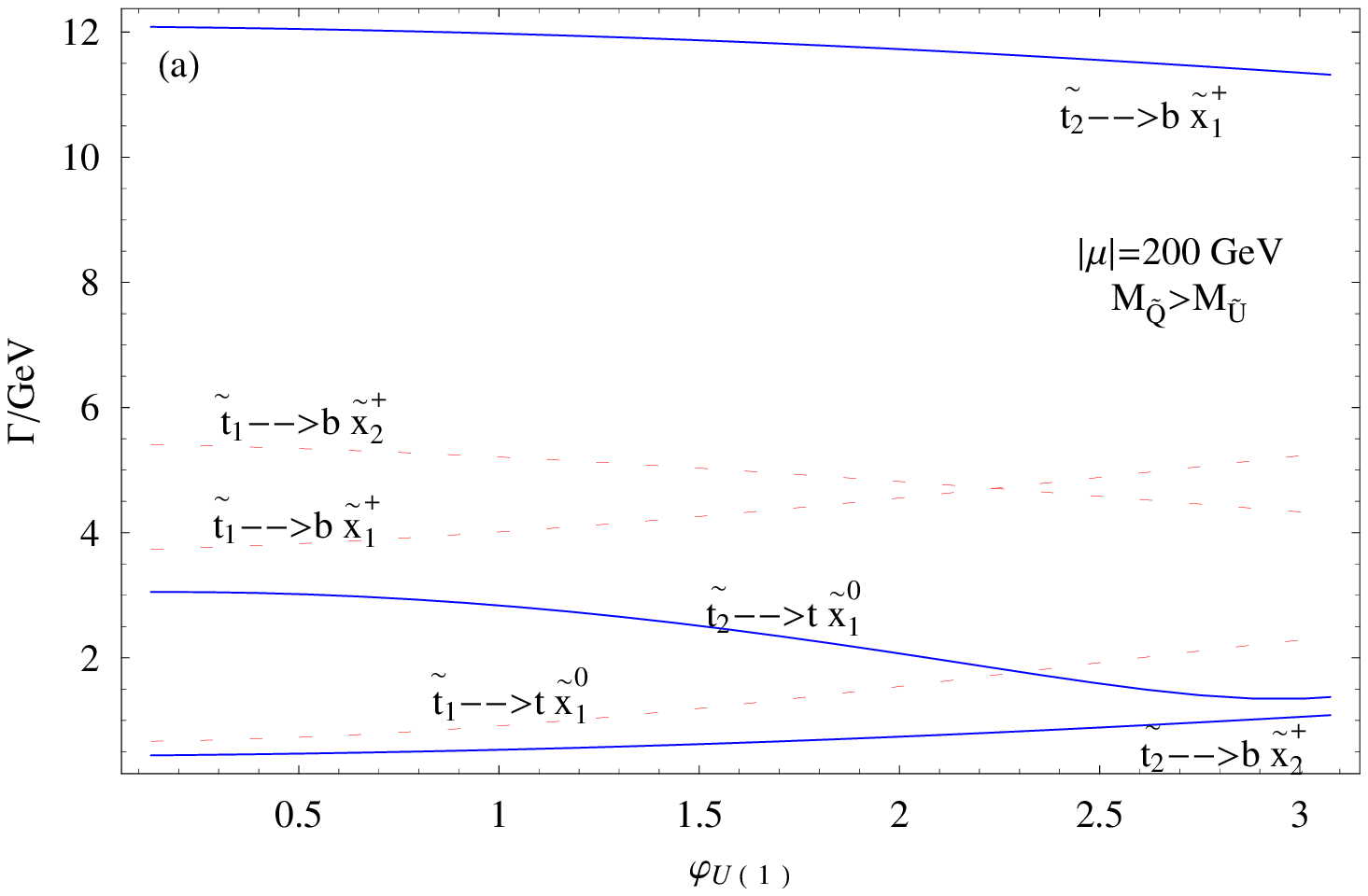} 
\label{figur5}
\end{figure}
\begin{figure}
\includegraphics{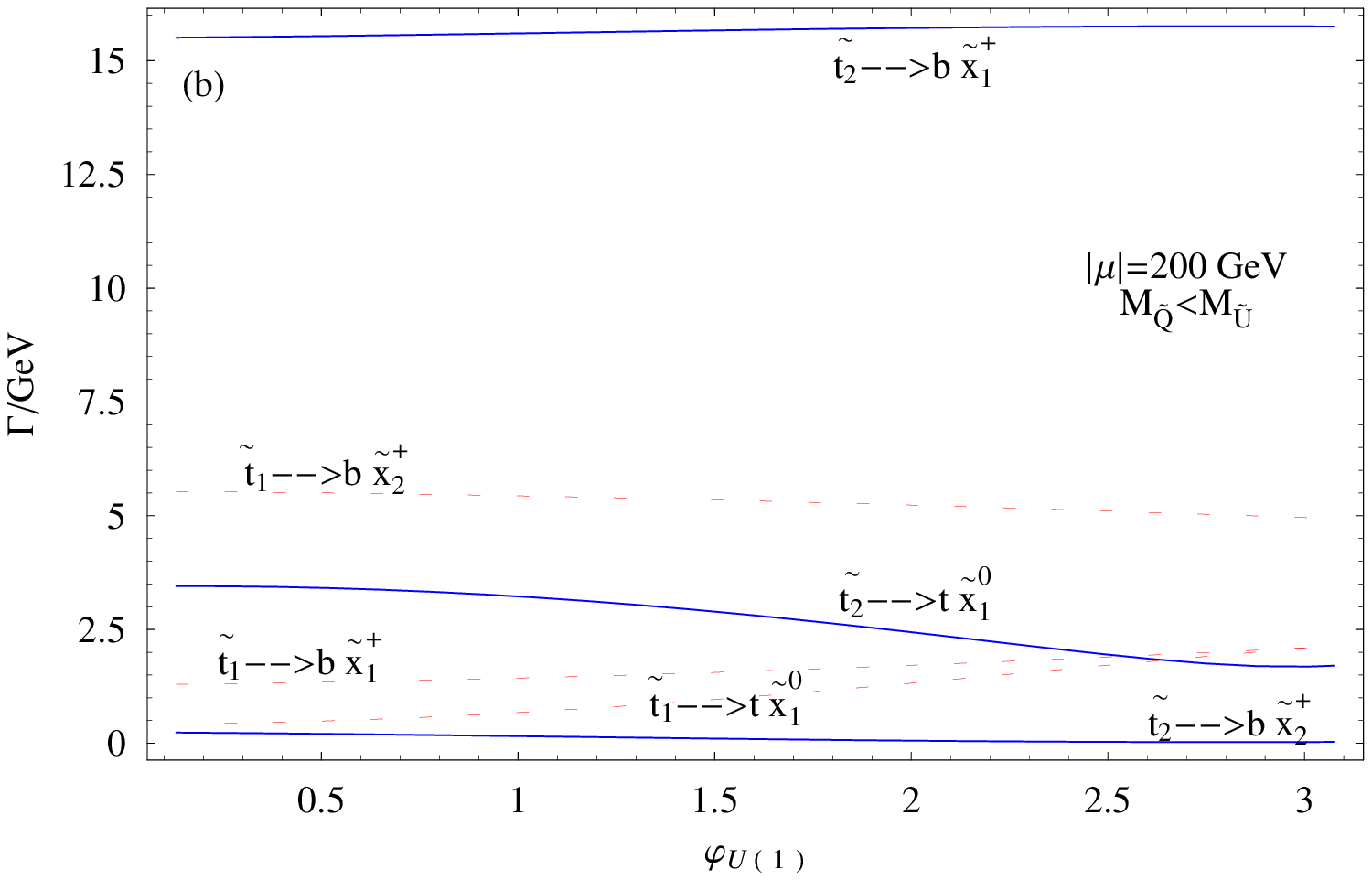} 
\caption{(a)-(b) Partial decay widths $\Gamma$ of the  $\tilde
t_{1,2}$ decays for $\mu =200 $ GeV , $\tan\beta=10$, $A_t$=1.2 TeV,
$\varphi_\mu$=$\varphi_{A_t}$=0,
  $m_{\tilde t_1}=750$ GeV and $m_{\tilde t_2}=1000$ GeV. } \label{figur6}
\end{figure}
\begin{figure}
\includegraphics{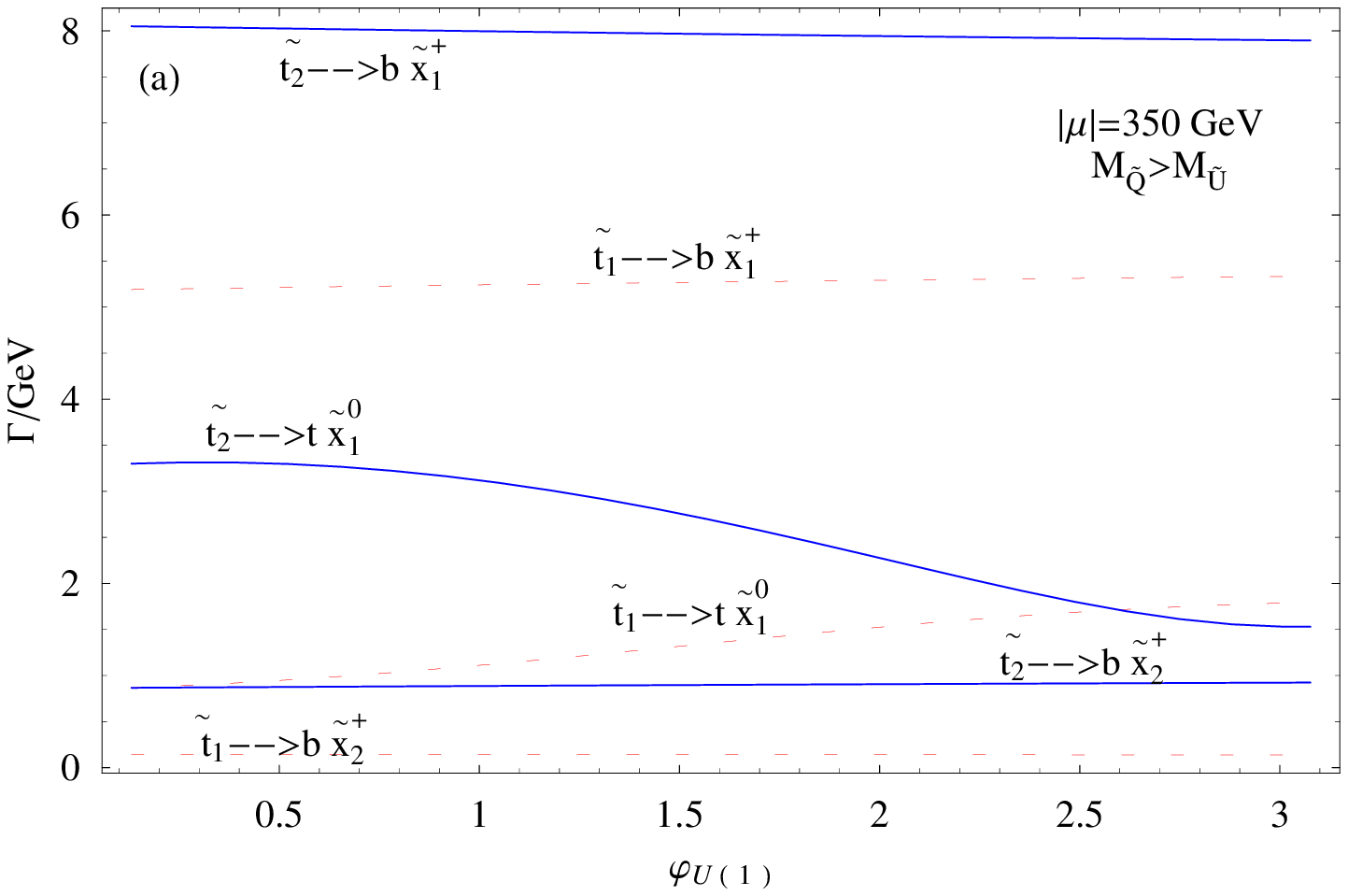} 
\label{figur7}
\end{figure}
\begin{figure}
\includegraphics{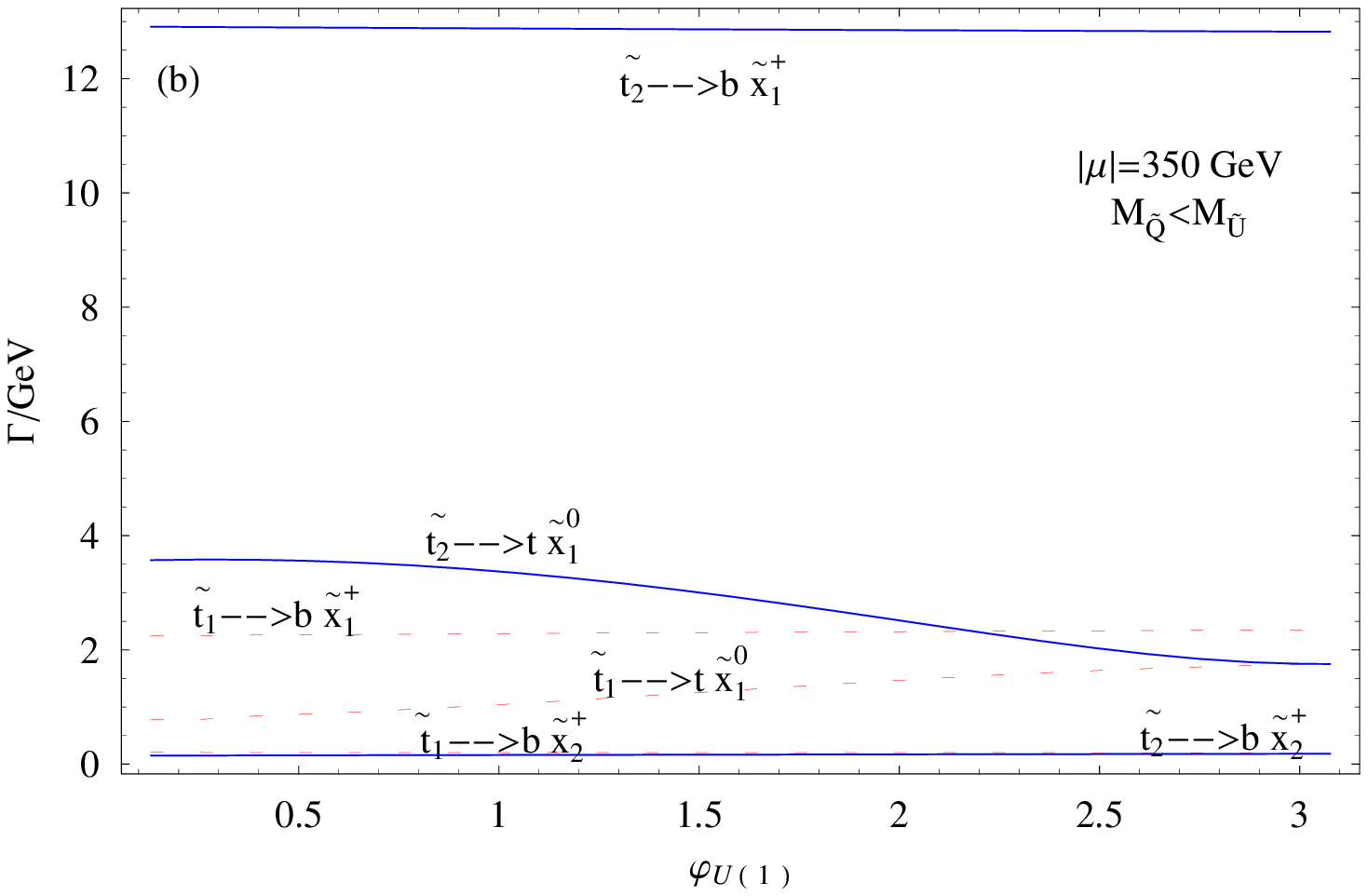} 
\caption{(a)-(b) Partial decay widths $\Gamma$ of the  $\tilde
t_{1,2}$ decays for $\mu =350 $ GeV , $\tan\beta=10$, $A_t$=1.2 TeV,
$\varphi_\mu$=$\varphi_{A_t}$=0,
  $m_{\tilde t_1}=750$ GeV and $m_{\tilde t_2}=1000$ GeV. }
 \label{figur8}
\end{figure}
\begin{figure}
\includegraphics{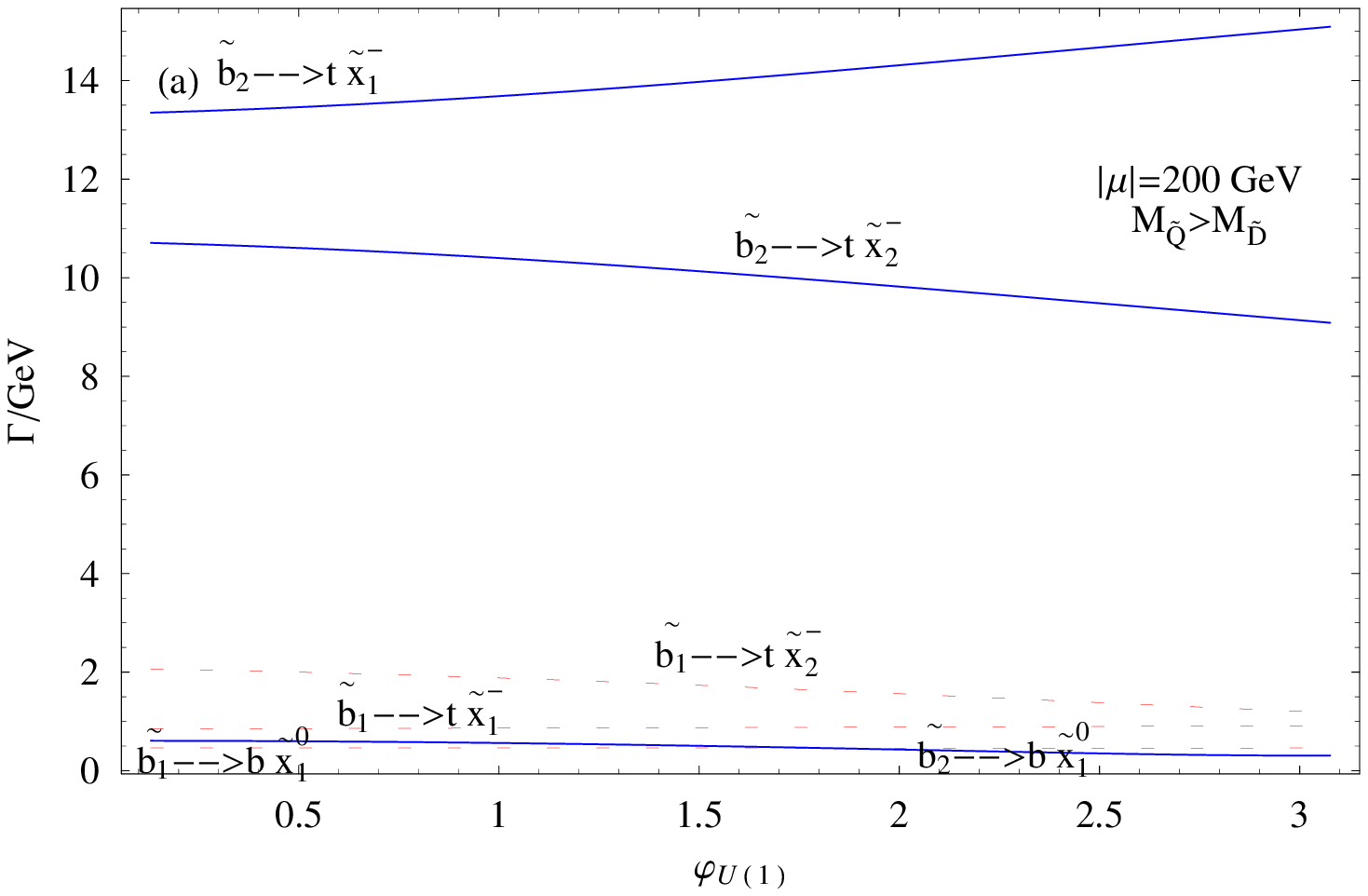} 
\label{figur7}
\end{figure}
\begin{figure}
\includegraphics{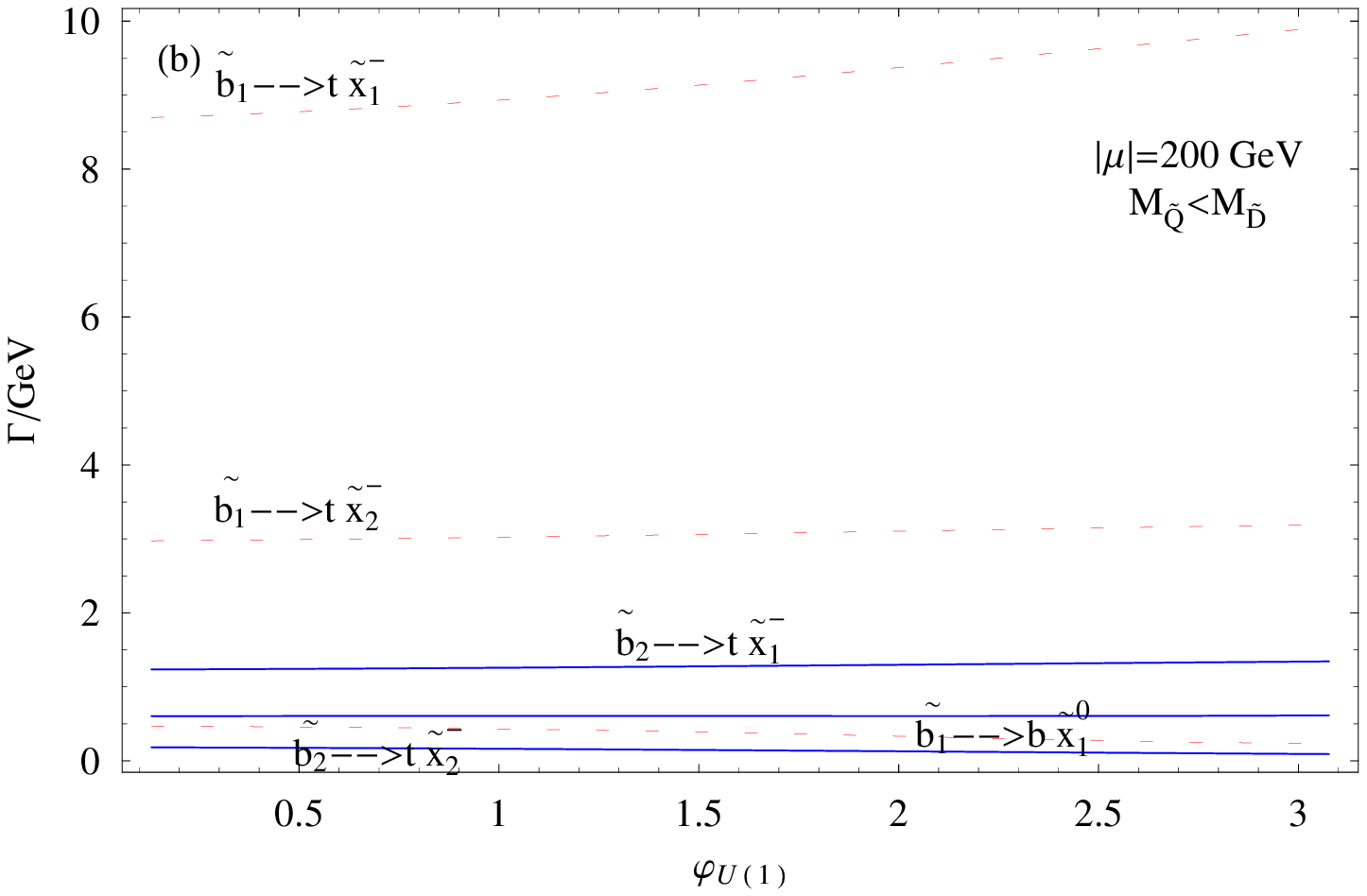} 
\caption{(a)-(b) Partial decay widths $\Gamma$ of the  $\tilde
b_{1,2}$ decays for $\mu =200 $ GeV , $\tan\beta=10$, $A_b$=1.2 TeV,
$\varphi_\mu$=$\varphi_{A_b}$=0,
  $m_{\tilde b_1}=750$ GeV and $m_{\tilde b_2}=1000$ GeV. }
 \label{figur8}
\end{figure}
\begin{figure}
\includegraphics{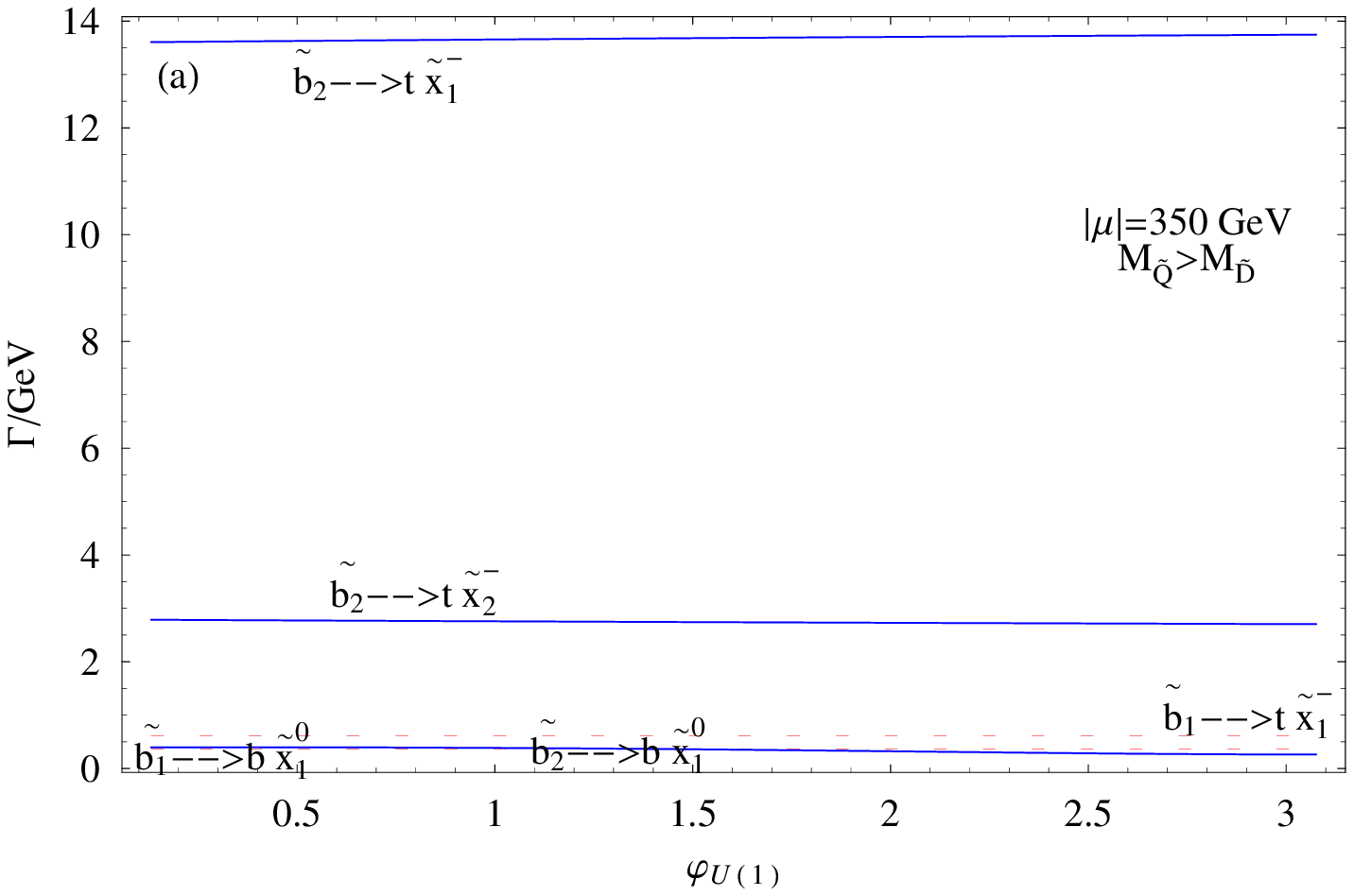} 
\label{figur7}
\end{figure}
\begin{figure}
\includegraphics{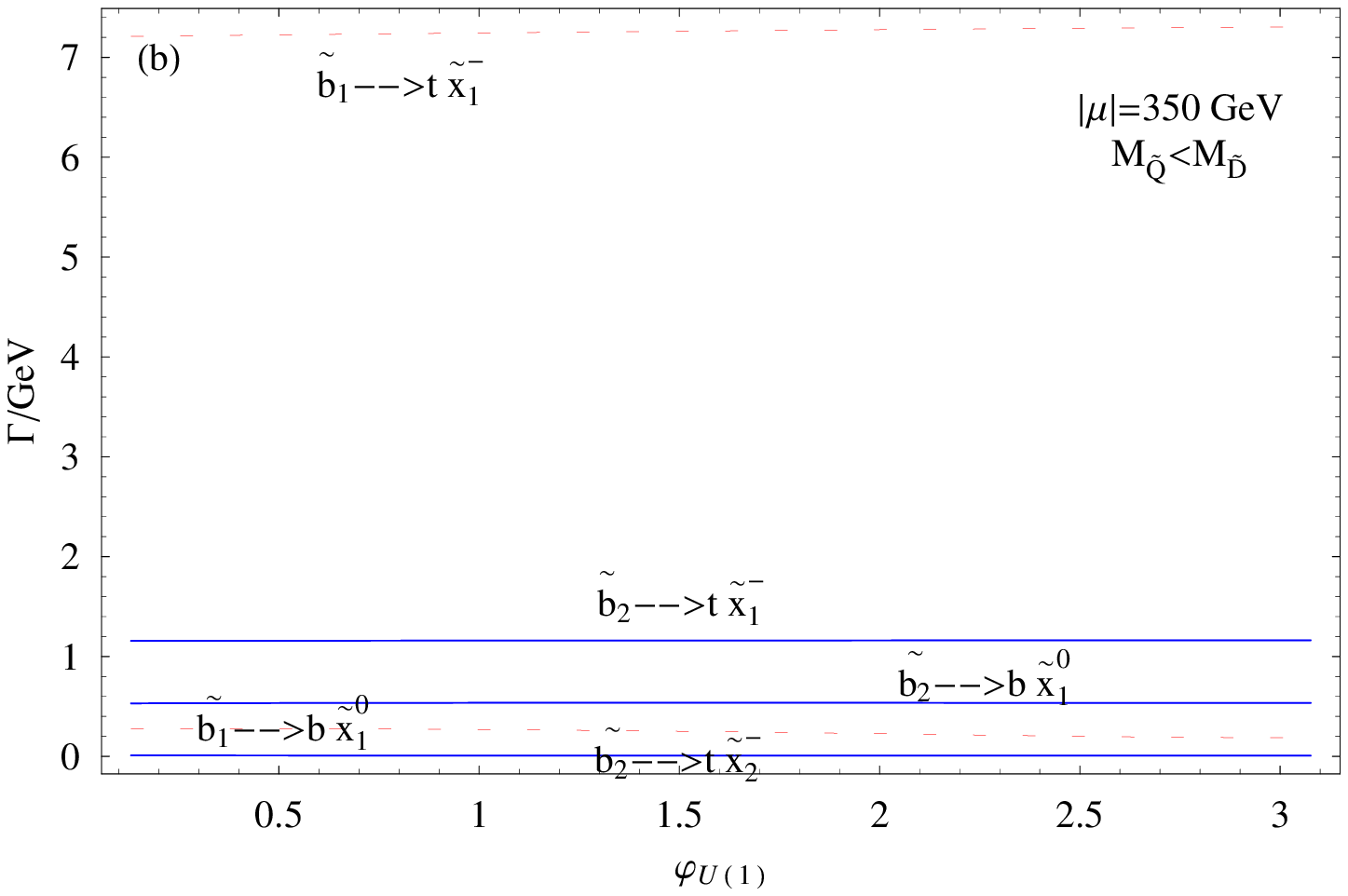} 
\caption{(a)-(b) Partial decay widths $\Gamma$ of the  $\tilde
b_{1,2}$ decays for $\mu =350 $ GeV , $\tan\beta=10$, $A_b$=1.2 TeV,
$\varphi_\mu$=$\varphi_{A_b}$=0,
  $m_{\tilde b_1}=750$ GeV and $m_{\tilde b_2}=1000$ GeV. }
 \label{figur8}
\end{figure}

\begin{thebibliography}{99}
\bibitem{haber}
H.~P.~Nilles, Phys.\ Rep.\ {\bf 110} (1984) 1; H.~E.~Haber and
G.~L.~Kane, Phys.\ Rep.\ {\bf 117} (1985) 75; R.~Barbieri, Riv.\
Nuovo Cim.\ {\bf 11} (1988) 1; J.~F.~Gunion and H.~E.~Haber, Nucl.\
Phys.\ B {\bf 272} (1986) 1 [Erratum-ibid. B {\bf 402} (1993) 567];
Nucl.\ Phys.\ B {\bf 278} (1986) 449
\bibitem{higgssector}
A.~Pilaftsis,
  Phys.\ Lett.\ B {\bf 435} (1998) 88
  [arXiv:hep-ph/9805373].
D.~A.~Demir,
  Phys.\ Lett.\ B {\bf 465} (1999) 177
  [arXiv:hep-ph/9809360];
  Phys.\ Rev.\ D {\bf 60} (1999) 055006
  [arXiv:hep-ph/9901389];
A.~Pilaftsis and C.~E.~M.~Wagner,
  Nucl.\ Phys.\ B {\bf 553} (1999) 3
  [arXiv:hep-ph/9902371].
\bibitem{Dugan:1984qf}
  M.~Dugan, B.~Grinstein and L.~J.~Hall,
  Nucl.\ Phys.\  B {\bf 255} (1985) 413;
  M.~J.~Duncan,
  Nucl.\ Phys.\  B {\bf 221} (1983) 285.
\bibitem{Masiero:2002xj}
  A.~Masiero and O.~Vives,
  New J.\ Phys.\  {\bf 4} (2002) 4.
\bibitem{Cabibbo:yz}
N.~Cabibbo,
Phys.\ Rev.\ Lett.\  {\bf 10}, 531 (1963);
M.~Kobayashi and T.~Maskawa,
Prog.\ Theor.\ Phys.\  {\bf 49}, 652 (1973).
\bibitem{bmeson}
D.~A.~Demir and K.~A.~Olive,
  Phys.\ Rev.\ D {\bf 65} (2002) 034007
  [arXiv:hep-ph/0107329];
P.~Gambino, U.~Haisch and M.~Misiak,
  Phys.\ Rev.\ Lett.\ {\bf 94} (2005) 061803
  [arXiv:hep-ph/0410155];
M.~E.~Gomez, T.~Ibrahim, P.~Nath and S.~Skadhauge,
  Phys.\ Rev.\ D {\bf 74} (2006) 015015
  [arXiv:hep-ph/0601163];
G.~Degrassi, P.~Gambino and P.~Slavich,
  Phys.\ Lett.\ B {\bf 635} (2006) 335
  [arXiv:hep-ph/0601135].
\bibitem{Goldberg:1983nd}
  H.~Goldberg,
  Phys.\ Rev.\ Lett.\  {\bf 50} (1983) 1419.
\bibitem{Ellis:1983ew}
  J.~R.~Ellis, J.~S.~Hagelin, D.~V.~Nanopoulos, K.~A.~Olive and M.~Srednicki,
  Nucl.\ Phys.\  B {\bf 238} (1984) 453.
\bibitem{Wilkinson}
D.~N.~Spergel  \textit{et al.} [WMAP Collaboration], Astrophys.\ J.\
Suppl {\bf 148} (2003)
175[arXiv:astro-ph/0302209];\\
C.~L.~Bennett \textit{et al.} [WMAP Collaboration], Astrophys.\ J.\
Suppl {\bf 148} (2003)1[arXiv:astro-ph/0302207];
\bibitem{Belanger}
G.~B\'{e}langer, F.~Boudjema, S.~Kraml, A.~Pukhov, A.~Semenov,
Phys.\ Rev.\ D {\bf 73} (2006) 115007[arXiv:hep-ph/0604150];
\bibitem{Spergel:2006hy}
  D.~N.~Spergel {\it et al.}  [WMAP Collaboration],
  Astrophys.\ J.\ Suppl.\  {\bf 170} (2007) 377
  [arXiv:astro-ph/0603449].
\bibitem{selbuz}
L.~Selbuz and Z.~Z.~Aydin,
 Phys.\ Lett.\ B {\bf 645} (2007) 228
  [arXiv:hep-ph/0612204];
\bibitem{edms}
D.~Chang, W.~Y.~Keung and A.~Pilaftsis,
  Phys.\ Rev.\ Lett.\ {\bf 82} (1999) 900
  [Erratum-ibid.\ {\bf 83} (1999) 3972]
  [arXiv:hep-ph/9811202];
S.~Abel, S.~Khalil and O.~Lebedev,
  Nucl.\ Phys.\ B {\bf 606} (2001) 151
  [arXiv:hep-ph/0103320];
D.~A.~Demir, M.~Pospelov and A.~Ritz,
  Phys.\ Rev.\ D {\bf 67} (2003) 015007
  [arXiv:hep-ph/0208257];
D.~A.~Demir, O.~Lebedev, K.~A.~Olive, M.~Pospelov and A.~Ritz,
  Nucl.\ Phys.\ B {\bf 680} (2004) 339
  [arXiv:hep-ph/0311314];
M.~Pospelov and A.~Ritz,
  Annals Phys.\ {\bf 318} (2005) 119
  [arXiv:hep-ph/0504231];
D.~A.~Demir and Y.~Farzan,
  JHEP {\bf 0510} (2005) 068
  [arXiv:hep-ph/0508236].
\bibitem{Ellis:1982tk}
  J.~R.~Ellis, S.~Ferrara and D.~V.~Nanopoulos,
  Phys.\ Lett.\  B {\bf 114} (1982) 231;
  W.~Buchmuller and D.~Wyler,
  Phys.\ Lett.\  B {\bf 121} (1983) 321;
  J.~Polchinski and M.~B.~Wise,
  Phys.\ Lett.\  B {\bf 125} (1983) 393;
  J.~M.~Gerard, W.~Grimus, A.~Masiero, D.~V.~Nanopoulos and A.~Raychaudhuri,
  Nucl.\ Phys.\  B {\bf 253} (1985) 93;
  P.~Nath,
  Phys.\ Rev.\ Lett.\  {\bf 66} (1991) 2565;
  Y.~Kizukuri and N.~Oshimo,
  Phys.\ Rev.\  D {\bf 45} (1992) 1806.
  Y.~Kizukuri and N.~Oshimo,
  Phys.\ Rev.\  D {\bf 46} (1992) 3025;
  T.~Falk and K.~A.~Olive,
  Phys.\ Lett.\  B {\bf 375} (1996) 196
  [arXiv:hep-ph/9602299].
\bibitem{Barger:2001nu}
  V.~D.~Barger, T.~Falk, T.~Han, J.~Jiang, T.~Li and T.~Plehn,
  Phys.\ Rev.\  D {\bf 64} (2001) 056007
  [arXiv:hep-ph/0101106].
\bibitem{Ibrahim:1997nc}
  T.~Ibrahim and P.~Nath,
  Phys.\ Lett.\  B {\bf 418} (1998) 98
  [arXiv:hep-ph/9707409];
  T.~Ibrahim and P.~Nath,
  Phys.\ Rev.\  D {\bf 57} (1998) 478
  [Erratum-ibid.\  D {\bf 58} (1998) 019901, D  {\bf 60} (1999) 079903, D  {\bf 60} (1999) 119901]
  [arXiv:hep-ph/9708456];
  T.~Ibrahim and P.~Nath,
  Phys.\ Rev.\  D {\bf 58} (1998) 111301
  [Erratum-ibid.\  D {\bf 60} (1999) 099902]
  [arXiv:hep-ph/9807501];
  T.~Ibrahim and P.~Nath,
  Phys.\ Rev.\  D {\bf 61} (2000) 093004
  [arXiv:hep-ph/9910553];
  M.~Brhlik, G.~J.~Good and G.~L.~Kane,
  Phys.\ Rev.\  D {\bf 59} (1999) 115004
  [arXiv:hep-ph/9810457];
  M.~Brhlik, L.~L.~Everett, G.~L.~Kane and J.~D.~Lykken,
  Phys.\ Rev.\ Lett.\  {\bf 83} (1999) 2124
  [arXiv:hep-ph/9905215];
  M.~Brhlik, L.~L.~Everett, G.~L.~Kane and J.~D.~Lykken,
  Phys.\ Rev.\  D {\bf 62} (2000) 035005
  [arXiv:hep-ph/9908326];
  A.~Bartl, T.~Gajdosik, W.~Porod, P.~Stockinger and H.~Stremnitzer,
  Phys.\ Rev.\  D {\bf 60} (1999) 073003
  [arXiv:hep-ph/9903402];
  A.~Bartl, T.~Gajdosik, E.~Lunghi, A.~Masiero, W.~Porod, H.~Stremnitzer and O.~Vives,
  Phys.\ Rev.\  D {\bf 64} (2001) 076009
  [arXiv:hep-ph/0103324];
S.~Abel, S.~Khalil and O.~Lebedev,
  Phys.\ Rev.\ Lett.\ {\bf 86}, 5850 (2001)
  [arXiv:hep-ph/0103031];
 S.~Abel, S.~Khalil and O.~Lebedev,
  Phys.\ Rev.\ Lett.\ {\bf 89}, 121601 (2002)
  [arXiv:hep-ph/0112260].
\bibitem{Ellis:1983ed}
  J.~R.~Ellis and S.~Rudaz,
  Phys.\ Lett.\  B {\bf 128} (1983) 248.
\bibitem{Gunion:1984yn}
  J.~F.~Gunion and H.~E.~Haber,
  Nucl.\ Phys.\  B {\bf 272} (1986) 1
  [Erratum-ibid.\  B {\bf 402} (1993) 567].
\bibitem{flavor}
J.~S.~Hagelin, S.~Kelley and T.~Tanaka,
  Nucl.\ Phys.\ B {\bf 415} (1994) 293;
D.~A.~Demir,
  Phys.\ Lett.\ B {\bf 571} (2003) 193
  [arXiv:hep-ph/0303249];
J.~Foster, K.~i.~Okumura and L.~Roszkowski,
  JHEP {\bf 0603} (2006) 044
  [arXiv:hep-ph/0510422].
\bibitem{Bartl}
A.~Bartl, S.~Hesselbach, K.~Hidaka, T.~Kernreiter, W.~Porod, Phys.\
Rev.\ D {\bf 70} (2004) 035003[arXiv:hep-ph/0311338];
\end{thebibliography}
\end{document}